\documentclass[12pt]{article}

\usepackage{graphicx}% Include  files

\title{Extremely rare interbreeding events can explain
Neanderthal DNA in modern humans}% Force line breaks with \\
\author
{Armando G. M. Neves$^{1}$ \\ Maurizio Serva$^{2}$\\
\\
\normalsize{$^{1}$Departamento de Matem\'atica, Universidade Federal de Minas Gerais,}\\
\normalsize{aneves@mat.ufmg.br}\\
\normalsize{Av. Ant\^onio Carlos 6627, 30123-970 Belo Horizonte - MG, Brazil}\\
\normalsize{serva@univaq.it}\\
\normalsize{$^{2}$Dipartimento di Matematica, Universit\`a dell'Aquila, I-67010 L'Aquila, Italy}\\
}%\collaboration{MUSO Collaboration}%\noaffiliation

\date{\today}% It is always \today, today,
\begin{document}
\maketitle
\begin{abstract}
Considering the recent experimental discovery of Green \textit{et al} that present 
day non-Africans have 1 to $4\%$ of their nuclear DNA of Neanderthal origin, we propose here 
a model which is able to quantify the interbreeding events between Africans and Neanderthals at the time they coexisted in the Middle East. 
The model consists of a solvable system of deterministic ordinary differential equations 
containing as a stochastic ingredient a realization of the neutral Wright-Fisher drift process. 
By simulating the stochastic part of the model we are able to apply it to the interbreeding of African and Neanderthal 
subpopulations and estimate the only parameter of the model, which is the number of individuals 
per generation exchanged between subpopulations. Our results indicate that the amount of 
Neanderthal DNA in non-Africans can be explained with maximum probability by the exchange of 
a single pair of individuals between the subpopulations at each 77 generations, but larger 
exchange frequencies are also allowed with sizeable probability. The results are compatible with a total interbreeding population of order $10^4$ individuals and with all living humans being descendents of Africans both for mitochondrial DNA and Y chromosome.
\end{abstract}

%\pacs{87.23.Kg, 87.10.-e, 87.10.Ca}% PACS, the Physics and Astronomy
                             % Classification Scheme.

%\tableofcontents

\section{Introduction}
The question of whether all of us, living humans, descend  exclusively from an anatomically 
modern African population which completely replaced archaic populations in other continents, 
or if Africans could have interbred with these local hominids has been the subject of a long 
lasting and interesting debate. The first of these possibilities, known as Out of Africa 
model, is based mainly on genetic evidence \cite{cann} further supported by paleontological 
\cite{stringer} and archaeological findings \cite{mcbreartybrooks}. The latter, known as 
Multiregional model, on the contrary, has been more supported by morphological studies 
\cite{thornewolpoff}, but recently it 
has also been found consistent with genetic data \cite{templeton2005}. 
A third, intermediate possibility, known as assimilation model \cite{fagundes2007}, 
suggests that Africans may have interbred with local archaic hominids to a limited extent.

The decision of which model correctly describes the origin of \textit{Homo sapiens} 
is obscured by the intricacies of the statistical methods proposed for evaluating the 
models themselves.
Examples of such intricate methods, their conflicting conclusions and subsequent debate 
are given in \cite{templeton2005,fagundes2007, templeton2010}.

In this paper we will describe by a simple and realistic model the dynamics of 
two subpopulations -- Africans and  Neanderthals -- interbreeding at a slow rate. 
In particular, we quantitatively determine the fequency of interbreeding events which are necessary in order that non-African living humans have between 1 and $4\%$ nuclear DNA of Neanderthal origin, according to the discovery of Green \textit{et al} \cite{greenetal}.
 
Among other important achievements, the recent seminal paper by Green \textit{et al} provides the first direct evidence of interbreeding of modern humans with archaic hominids, Neanderthals in this case. By direct evidence we mean having sequenced Neanderthal nuclear DNA and showing that this DNA 
is more similar to nuclear DNA of living non-Africans than to nuclear DNA of living Africans.

Of course, the findings of Green \textit{et al} await anxiously for replication by the 
scientific community. Improvements in the resolution of the genome sequencing, in the comparison with present day individuals and DNA sequencing of other fossils classified as 
Neanderthals, \textit{H. erectus},  \textit{H. floresiensis} and modern humans are mostly welcome. 

Based on their findings and on archaeological evidence \cite{baryosef}, it was 
suggested in \cite{greenetal} that interbreeding between anatomically modern Africans 
and Neanderthals might have occurred in the Middle East before expansion of modern Africans 
into Eurasia, at a time in which both coexisted there. This hypothesis is assumed in this paper, allowing inference of the only parameter in the model, 
the rate of exchange of individuals between Africans and Neanderthals, and giving some idea on the size of the total population involved in the interbreeding. 

The model will be fully explained in the next section, but we anticipate here its main features. Total population size is supposed fixed, but 
African and Neanderthal subpopulations sizes fluctuate according to the neutral (i.e. Africans and Neanderthals are supposed to have the same fitness) Wright-Fisher model \cite{ewens} for two alleles at a single locus. We also assume no biological barriers for interbreeding and no strong hypotheses on the initial composition of the population. 
Gene flow between subpopulations is implemented by assuming that a fixed number $\alpha$ of pairs of individuals per generation is exchanged between them.

The model is characterized by a 
deterministic component -- a system of two linear ordinary differential equations (ODEs) -- and a stochastic component -- a realization of the Wright-Fisher drift process to be 
introduced as an external function in the ODEs.
The ODEs are exactly solvable, up to definite integrals depending on the stochastic part.
The stochastic part can be dealt with by simple simulations.

Assuming a random initial fraction of Africans, our main result is the conditional probability density distribution for the exchange parameter $\alpha$, illustrated in Fig. \ref{novafig}. The condition to be satisifed is that, after interbreeding with Neanderthals,  a fraction of 1 to $4 \%$ of Neanderthal genes, as suggested by \cite{greenetal}, will be present in the African population.  Fig. \ref{novafig} shows this condition is attained with maximum probability
for $\alpha_{\mathrm{max}} \approx 0.013$, i.e. one pair of individuals is 
exchanged between the two subpopulations every 77 generations. The mean value of $\alpha$ is $\alpha_{\mathrm{mean}} \approx 0.083$, which corresponds to one pair of individuals exchanged at each 12 generations.

Such conclusions are based on a solvable mathematical model and simple simulations,
avoiding statistical in favor of probabilistic methods. 
Application of probabilistic methods reminiscent of Statistical Mechanics to biological problems has been abundant in the literature of Physics and Mathematics 
communities, but penetration into 
Biology and Anthropology has proved more difficult. 
In particular, both authors of this paper have previously and separately anticipated
\cite{nm1,nm2,nm3,serva1,serva2,serva3} that 
evidences based on mitochondrial DNA (mtDNA) could not rule out the possibility of 
interbreeding among modern humans and other archaic forms. We hope that the direct 
experimental proof of such interbreeding provided by \cite{greenetal} can be the occasion 
for better acceptance of methods such as the ones we will discuss.

While writing the present paper a new report \cite{reichetal} concerning the
interbreeding of modern humans with another archaic hominid group was published. 
Results have been obtained by studying the fossil nuclear DNA extracted from  
the finger of a single individual previously known only from its mtDNA  \cite{krauseetal}.
The individual is considered a representative
of an archaic group of hominids (Denisovans) different both from moderns and Neanderthals. 
According to the authors, Denisovan nuclear DNA is present in living Melanesians 
in a proportion of about $6 \%$. Very few is known about the morphology of Denisovans, 
as complete fossils belonging to this group are not yet known.

Although we still have no data concerning
the size of populations and the duration of coexistence, the model described in this paper might be used 
to describe the interbreeding between modern humans and Denisovans.

\section{The model}

Consider a population of constant size equal to $N$ individuals. 
We suppose that the population is divided into two subpopulations we call 1 and 2, 
generations are non-overlapping and the number of generations is counted from past to future. 
Reproduction is sexual and diploid. We also suppose that the subpopulations have lived isolated from each other 
for a long time before they meet. At generation $g=0$, when subpopulations meet, 
the total population consists then of two groups, each of which consisting in individuals 
of a pure race. Starting at this time subpopulations will share a common environment for a 
long period.

We do not suppose that the numbers $N_1(g)$ and $N_2(g)$ of individuals at generation 
$g$ in each of the two subpopulations are constant, although their sum $N_1(g)+N_2(g)=N$ is. 
Instead, $N_1(g+1)$ and $N_2(g+1)$ are random variables which can be determined by 
Wright-Fisher rule i.e., any of the $N$ individuals of generation $g+1$ independently chooses 
to belong to subpopulation 1 with probability 
$N_1(g)/N$ and to subpopulation 2 with probability  $N_2(g)/N$. 
After that, both father and mother of an individual in generation $g+1$ are uniformly 
randomly chosen among all males and females of generation $g$ in the subpopulation he/she 
has chosen. 

With such a reproduction mechanism the numbers $N_1(g)$ and $N_2(g)$ 
fluctuate as generations pass until one of the subpopulations becomes extinct. 
This stochastic process is the same as in the simplest version of the 
neutral, i.e. no selective advantage for any of the alleles,
Wright-Fisher model for two alleles at a single locus \cite{ewens}. 
The time for extinction is random as well as which of the two subpopulations becomes extinct. 
If $x(0)= N_1(0)/N$ is the initial fraction of individuals of subpopulation 1, 
then subpopulation 1 will survive with probability $x(0)$ and the mean number of generations 
until extinction is 
$-2N [x(0) \ln x(0) + (1-x(0)) \ln (1-x(0))]$ (see \cite{ewens}). 
As the mean number of generations for extinction of one subpopulation scales with $N$, 
it is reasonable to measure time not in generation units, 
but in generations divided by $N$. From here on, we will refer to $t=g/N$ simply 
as \textit{time} and we will refer to $x(t)=N_1(t)/N$ in a realization of the above 
stochastic process as the \textit{history} of the Wright-Fisher drift process.

In the previously described dynamics no mechanism of gene admixture between 
subpopulations was present and we add it as follows.
We assume that at each generation a number $\alpha$ of random individuals from 
subpopulation 1 migrates to subpopulation 2 and vice-versa the same number of random
individuals from subpopulation 2 migrates to subpopulation 1.
In other words, $\alpha$ \textit{pairs} per generation are exchanged. We strongly underline that $\alpha$
is a number of order 1, not of order $N$. Migrants will contribute with their genes for the next generation just like any other 
individual in their host subpopulation. 
Their offspring, if any, is considered as normal members of the host subpopulation. 

The parameter $\alpha$ introduced above may be non-integer and also less than 1. 
In such cases we interpret it as the average number of pairs of exchanged individuals per generation.

By the hypothesis of isolation between subpopulations for a long time before $t=0$, 
we may suppose that in many {\it loci} the two subpopulations will
have different and characteristic alleles.
Therefore, we can assume that there exists a large set of alleles which are exclusive of 
subpopulation 1 and the same for subpopulation 2. 
We will refer to these alleles respectively 
as \textit{type 1} and \textit{type 2}.
At any time $t\geq0$ any individual will be characterized by his/her fractions of type 1 
and type 2 alleles. 
We define then $y_1(t)$ as the \textit{mean fraction of type 1 alleles in subpopulation 1 
at time $t$} and $y_2(t)$ \textit{as the mean fraction of type 1 alleles in subpopulation 2 
at time $t$}. 
The \textit{mean} here is due to the fact that individuals in subpopulation 1 
in general have different allelic fractions, but $y_1(t)$ is calculated by averaging 
allelic fractions among all individuals in subpopulation 1, and 
similarly for $y_2(t)$. Of course $y_1(0)=1$ and $y_2(0)=0$. 
Similar quantities might have been defined for type 2 alleles, 
but they are easily related to $y_1(t)$ and $y_2(t)$ and thus unnecessary.

It is now possible to derive the basic equations relating the mean allelic fractions at 
generation $g+1$ with the mean allelic fractions at generation $g$. In doing so we will 
make the assumption that the $\alpha$ individuals of subpopulation 1 migrating to 
subpopulation 2 all have an allelic fraction equal to $y_1(t)$. 
The analogous assumption will be made for all the individuals of subpopulation 2 
migrating to subpopulation 1.

Of course the above assumption of exchanged individuals
all having the mean allelic fractions in their subpopulations is 
a very strong one and it is not strictly true. Nonetheless, it is indeed a very good 
approximation if $\alpha$ is much smaller than $1/\log_2 N$. In fact, $1/\alpha$ is the 
number of generations between two consecutive exchanges of individuals. As the 
typical number of generations for genetic homogenization in a population of $N$ individuals 
with diploid reproduction and random mating is $\log_2 N$, 
see \cite{derrida1,derrida2,derrida3,chang}, the  condition that $\alpha$ is much 
smaller than $1/\log_2 N$ makes sure that subpopulations 1 and 2 are both rather homogeneous 
at the exchange times. 

The allelic fraction $y_1(t+1/N)$ will be equal to $y_1(t)$ plus the contribution of type 1 
alleles from the immigrating individuals of subpopulation 2 and minus the loss of type 1 
alleles due to emigration. We remind that these loss and gain terms are both proportional 
to $\alpha$ and inversely proportional to the number $N x(t)$ of individuals in 
subpopulation 1. Similar considerations apply to $y_2(t+1/N)$.
In symbols:
\begin{equation}  \label{eqdif}
\left\{\begin{array}{rcl}
y_1(t+\frac{1}{N}) &=& \left(1- \frac{\alpha}{N x(t)}\right) 
\,y_1(t) \,+\, \frac{\alpha}{N x(t)} \, y_2(t) \\
y_2(t+\frac{1}{N}) &=& \frac{\alpha}{N (1-x(t))} \, y_1(t) \,+\,
\left(1- \frac{\alpha}{N (1-x(t))}\right) \,y_2(t) \end{array}\right. \;.
\end{equation}

The above equations, after taking the $N \rightarrow \infty$ limit,
become a system of linear ODEs
\begin{equation}  \label{odes}
\left\{\begin{array}{rcl}
y_1'(t) &=& - \frac{\alpha}{x(t)} \, (y_1(t)-y_2(t)) \\
y_2'(t) &=& \frac{\alpha}{1-x(t)} \, (y_1(t)-y_2(t)) \end{array}\right. \;.
\end{equation}

We stress here that we think of $x(t)$ as a stochastic function obtained by realizing 
the Wright-Fisher drift, but Eqs. (\ref{eqdif})  and (\ref{odes}) still 
hold if $x(t)$ is any description of the history of the size of subpopulation 1, 
be it stochastic or deterministic. For example, the possibility of individuals in subpopulation 1 being fitter than individuals in subpopulation 2 has been explored, still using (\ref{eqdif}), in another work \cite{biomat2011}.

Eqs. (\ref{odes}) can be exactly solved up to integrals depending on $x(t)$. Although such integrals cannot be calculated in general, the exact solution can be used to give a qualitative view of the behaviour of functions $y_1(t)$ and $y_2(t)$. It turns out that $y_1$ is a decreasing function, whereas $y_2$ increases. The decrease and increase rates are larger when $\alpha$ is large and, despite symmetry in our immigration assumption, gene flow between subpopulations is in general asymmetrical. Such features are shown in appendix A.

Moreover, Eqs. (\ref{eqdif}) lend themselves to simple and rapid numerical solution for quantitative purposes. In appendix B, we address the question of comparing numerical solutions of Eqs. (\ref{eqdif}) and direct simulation of all stochastic processes involved. We see that there is good agreement between simulations and numerical solutions of Eqs. (\ref{eqdif}). In all that follows, unless explicitly stated, we will use results obtained by numerically solving Eqs. (\ref{eqdif}), because the computer time for numerical solution is much smaller than for simulation.

\section{Estimating the exchange parameter}

We know that Neanderthals were extinct and, according to \cite{greenetal}, before disappearing 
they interbred with modern humans. Despite comparisons between nuclear DNA of Neanderthals and 
living humans having been limited up to now by a sample of only 3 Neanderthals and 5 living humans, 
the authors of \cite{greenetal} observed that all three non-Africans in their sample are equally 
closer to the Neanderthals than the two Africans. They estimate that non-African living humans 
possess $1$ to $4\%$ of their nuclear DNA derived from Neanderthals. 
Supposing that Africans are subpopulation 1 in our model, this means that the final value 
of $y_1$ should lie between $0.96$ and $0.99$ in order to comply with their experimental 
conclusions. We will refer in the following to the interval between 0.96 and 0.99 as the 
\textit{experimental interval} for the final value of $y_1$.

As we do not know the composition of the total population at the time the two subpopulations met, 
we will take the initial fraction $x(0)$ of Africans as a random number.
With this hypothesis, the only free 
parameter is the exchange rate $\alpha$.

As can be seen in Fig. S1 the value of $\alpha$ largely influences the 
final value of $y_1$. Furthermore, in both
Figs. S1 and S2 it can be seen that with 
$\alpha=1$ or $\alpha=0.1$ the final values of $y_1$ tend to be too small to be compatible 
with the experimental interval. 
We stress that these figures are based only on 
two realizations of the history $x(t)$ and a single value $x(0)=0.5$. 
In order to produce estimates of $\alpha$ we must produce a large number of histories $x(t)$ 
with many values of $x(0)$ and for any of these simulated histories recursively solve 
Eqs. (\ref{eqdif}) in order to determine the associated final value of $y_1$.

\begin{figure*}
	\centering
		\includegraphics[width=0.9\textwidth]{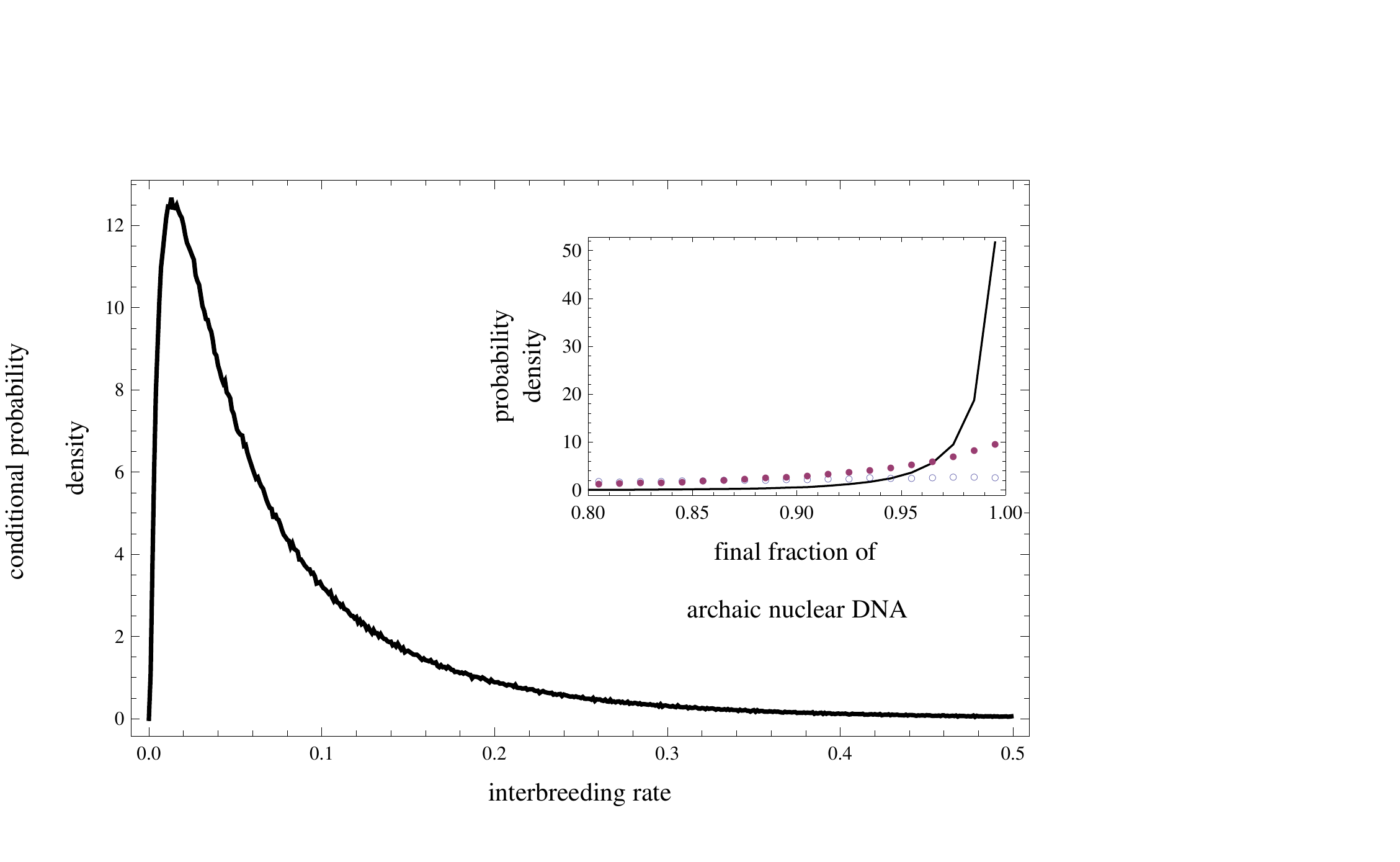}
			%\begin{picture}(21, 0.3)
		%\put(4.4,0){(a)}
		%\put(13,0){(b)}
		%\end{picture}
	\caption{\label{novafig} The $\alpha$ probability density, i.e. the probability 
density that the final value of $y_1$  is in the experimental interval 
0.96 - 0.99 given a value of $\alpha$.  The plot was built by obtaining one million ``successful" pairs $(x(t),\alpha)$ such that the final value of $y_1$ obtained by solving Eqs. (\ref{eqdif}) lies in the experimental interval. These pairs were obtained out of a total of around 140 million simulations with random $x(0)$ uniformly distributed between 0 and 0.8 and $\alpha$ uniformly distributed between 0 and 2. For the sucessful pairs we then computed
the fraction associated to any given  $\alpha$. In the inset we plot the probability 
density for the theoretical final values of $y_1$ for three  different values of $\alpha$.
The densities are empirically determined by simulating 400,000 
Wright-Fisher drift histories $x(t)$ with random $x(0)$ uniformly distributed between 0 
and 1 and selecting the histories in which 
subpopulation 2 is extinct. The empty dots (blue) are data for $\alpha=1$, the full dots (purple)
are data for $\alpha=0.1$ and the full curve (black) are for $\alpha=0.01$.}
\end{figure*}

The inset in Fig. \ref{novafig} is realized by producing 400,000 Wright-Fisher drift histories
$x(t)$ with random $x(0)$ uniformly distributed between 0 and 1.  For all these histories 
we compute the final theoretical value of $y_1$ 
by solving Eqs. (\ref{eqdif}) using the three values $\alpha=1$, $\alpha=0.1$ and $\alpha=0.01$.
Therefore, for each of the three values of $\alpha$ we have about 200,000 data which 
allow inference of the probability density for the final value of $y_1$. 
The data plotted in the inset of Fig. \ref{novafig} show
that for $\alpha=1$ the probability that the final value of $y_1$ lies in 
the experimental interval is approximately equal to $8.1\%$. For $\alpha=0.1$ the corresponding 
probability is approximately of  $21.5\%$ and for $\alpha=0.01$ it is approximately of  $34.0\%$. 
In all three cases the density of the final values of $y_1$ is rather thick, meaning that 
there is a large probability that the final value of $y_1$ 
does not lie in the experimental interval.

The above information shows that the experimental data are better explained 
by values of $\alpha$ much smaller than 1. 
By Fig. \ref{novafig} we see that the value of $\alpha$ 
which explains with largest probability the experimental data is 
$\alpha_{\mathrm{max}} \approx 0.013$. 
In order to produce that plot, we simulated  a large number of Wright-Fisher histories 
$x(t)$ with random $x(0)$ uniformly extracted between 0 and 0.8 and random values 
for $\alpha$ uniformly distributed between 0 and 2. 
From these data we selected the histories in which subpopulation 2 was extinct and such 
that the final theoretical value of $y_1$ lied in the experimental interval. 
In this way we can empirically determine the probability 
that the final value of $y_1$ lies in the experimental interval as a function of $\alpha$ .

We also see that the probability density for $\alpha$ is rather asymmetrical 
around $\alpha_{\mathrm{max}}$ with 
values $\alpha \ge\alpha_{\mathrm{max}}$ contributing with large probability. This asymmetry
is reflected in the fact that the mean value is  $\alpha_{\mathrm{mean}} \approx 0.083$, 
much larger than $\alpha_{\mathrm{max}}$. 

A technical detail in producing Fig.\ref{novafig} is that the random values for $x(0)$
are chosen with uniform distribution in the interval $0 \le x(0)\leq 0.8$, avoiding 
values either close to or inside the experimental interval. Such a choice is related to the assumption of \textit{slow} rather than \textit{rapid} interbreeding between Africans and Neanderthals. See aapendix C and Fig. S3 for a more detailed explanation on that choice.

\section{Other results}
O. Bar-Yosef \cite{natgeo} compares occupation of the Middle East by Neanderthals and Africans with a long football game. The occupants of the caves of Skuhl and Kafzeh in Israel alternated between Africans and Neanderthals several times over a period of more than 130,000 years. Although the model described before becomes independent of the total population $N$, we may obtain some hints on the size of $N$ if we accept the constraint that at least for 130,000 years Neanderthals had not been extinct in the Middle East.

By taking random values for $x(0)$ between 0 and 0.8 and $\alpha$ between 0 and 2 we obtained a sample of 790 events such that Neanderthals were extinct and $y_1$ lied in the experimental interval. For each of these events we recorded the time it took for extinction of Neanderthals and we found out that the mean extinction time was 0.58. If we take this mean value as the typical value, suppose that one generation is 20 years and equate it to 130,000 years, we get $N \approx 11,200$ individuals. The whole distribution of extinction times in the above sample is shown in Fig. S4.

In Fig. S3 we plotted the same sample of events in the plane $x(0) - \alpha$. We see that smaller values of $x(0)$ are correlated with smaller values of $\alpha$ and also that the events such that $y_1$ lies in the experimental interval are concentrated around the largest values of $x(0)$. The mean value of $x(0)$ for the whole sample is 0.64.

Using the same sample we may also explore the values of $y_2$ at the time Neanderthals were extinct, i.e. the fraction of African DNA in the last Neanderthals which interbred with Africans. Fig. S5 shows a histogram of the $y_2$ values for the events in the sample. Observe that typical values of $y_2$ are much larger than the values of $1-y_1$, which range from 0.01 to 0.04. This is due to the fact that in most events such that $y_1$ falls within the experimental interval, Africans were the majority of population for most of the time. According to the explanation in appendix A, this implies that, despite symmetry in the number of exchanged individuals, transfer of African alleles to Neanderthals will be larger than transfer of Neanderthal alleles to Africans.

By simulating the complete reproduction and individual exchange process described in appendix B we were able also of empirically determining the conditional probability -- the condition being that the fraction of African DNA in Africans is in the experimental interval -- that the most recent common ancestors in the population for
the maternal (mtDNA) and paternal lineages (Y chromosome) are both African. 
We ran several simulations with populations of 100 individuals and random values of $\alpha$ 
uniformly distributed between 0.01 and 0.2 and random $x(0)$ 
constrained to be smaller than 0.8. In each simulation we waited until all male individuals had 
the same paternal ancestor and all female individuals had the same maternal ancestor. 
We selected those simulations in which subpopulation 1 survived and $y_1$ lied in the experimental 
interval. Out of 96 simulations satisfying the above criteria, only in 7 of them 
the survived Y chromosome and mtDNA lineages were not both of ancestors belonging to subpopulation 1. Therefore, according to our interbreeding model, the conditional probability of an African origin
of both mtDNA and Y chromosome can be estimated to be of order $0.93$.

\section{Discussion and conclusions}
Large samples of mtDNA \cite{cann} and Y chromosomes 
\cite{underhill} in living humans have been sequenced. 
The small variation among living humans is compatible with a single ancestor woman (mtDNA) and 
a single ancestor man (Y chromosome) to the whole population, probably both of African origin and living about 100-200 
thousand years ago. These facts have been interpreted as proofs of the Out of Africa model, but our interbreeding model is perfectly compatible with them. In fact, conditioned to $y_1$ being in the experimental interval, our model yields a large probability of 93\% for African origin of both mtDNA and Y chromosome. 

More recently \cite{kringsetal}, the whole mtDNA of a few Neanderthal fossils became available. 
The average number of pairwise differences in mtDNA between a Neanderthal and a living human is 
significantly larger than the average number of pairwise differences in mtDNA among living humans. 
This has been considered as a further confirmation of the claim that Neanderthals belong to a 
separate species, see e.g. \cite{curratexcoffier}, and also for the Out of Africa model.

Before any data on Neanderthal nuclear DNA was available, both authors of this paper had separately anticipated \cite{serva1,serva2,serva3,nm1,nm2,nm3} that the above facts are all compatible 
with anatomically modern Africans and Neanderthals being part of a single interbreeding 
population at the times they coexisted. Some further details about these claims are given in appendix E.

In the framework of the model proposed in this article we could infer that the 1 to 4\% fraction \cite{greenetal}
of Neanderthal DNA in present day non-Africans can be explained with maximum 
probability by assuming that the African and Neanderthal subpopulations exchanged 
only 1 pair of individuals in about 77 generations. But the mean value of the exchange 
parameter in the model corresponds to a larger frequency of about 1 pair of individuals 
exchanged in about 12 generations. 

We also estimated the mean number of generations for Neanderthal extinction in the Middle East to be approximately $0.58 N$. Together with the fact that Neanderthals and Africans seem to have coexisted in the Middle East for at least 130,000 years, this allows us to estimate the total population $N$ in the model to be of order $10^4$ individuals.

Although Green \textit{et al} have observed in \cite{greenetal} gene flow from Neanderthals into Africans, they have not observed the reverse flow. This fact is also compatible both with our results and the fact that living Europeans are as close to Neanderthals as living Asians or Oceanians. The explanation is that the Neanderthal specimens which had their DNA sequenced in \cite{greenetal} were all excavated in European sites. It seems that only a part of the total Neanderthal population took part in the interbreeding process in Middle East, the other part of the population remaining in Europe. The descendants of these Neanderthals which have never left Europe did not interbreed later with Africans when they came into Europe, or this interbreeding was very small. On the contrary, according to our model, see Fig. S5, we expect to find a larger fraction of African DNA in late Middle East Neanderthal fossils than the 1 to 4\% Neanderthal fraction of present non-Africans. Thus, DNA sequencing of one such fossil would be a good test for the present model.

Neanderthals are implicitly considered in this work as a group within the 
\textit{Homo sapiens} species and we renounce the strict Out of Africa model for the 
origin of our species, in which anatomically modern Africans would have replaced without 
gene flow other hominids in Eurasia. In particular, our model is neutral in the sense that 
we assign the same fitness to Neanderthals and Africans. Our results show that neither strong 
sexual isolation between Africans and Neanderthals or else some kind of Neanderthal cognitive 
or reproductive inferiority, are necessary to explain both their extinction and the small 
fraction of their DNA in most living humans. 
In fact, within the assumptions of the model, if two subpopulations coexist in the same 
territory for a sufficiently long time, only one of them survives.  
The fact that Neanderthals were the extinct subpopulation is then a random event.

Although we do not intend to back up any kind of superiority for Neanderthals, our neutrality hypothesis is at least supported by
recent results \cite{zilhaoetal,wongzilhao} by J. Zilh\~ao \textit{et al}, which claim that Neanderthals 
in Europe already made use of symbolic thinking before Africans arrived there.

Current knowledge about Denisovans morphology and life style is much less than what 
we know about Neanderthals. In particular we do not know whether Denisovans lived only in
Siberia, where up to now the only known fossils have been found, or elsewhere. 
Where and when this people made contact with the
African ancestors of present day Melanesians is still a mystery. 
Nevertheless, if such a contact occurred for a sufficiently long time in a small 
geographical region, then the present model can be straightforwardly applied.

As we now know of our Neanderthal and Denisovan inheritances, it is time to ask whether 
they were the only hominids that Africans mated. We believe that the future may still uncover
lots of surprises when Denisovans will be better studied and nuclear DNA of many more Neanderthal and other hominid fossils will become available. 

%\bibliographystyle{plain}
%\bibliography{neandertalDNA}

\appendix

\renewcommand{\thefigure}{S\arabic{figure}}   %%label figures as S1, S2, and so on

\renewcommand{\theequation}{S\arabic{equation}} 

\setcounter{equation}{0}
\setcounter{figure}{0}

\section{Solution and qualitative behaviour of solutions of the model equations}
By introducing the auxiliary functions $z_1(t)=y_1(t)-y_2(t)$ and
$z_2(t)=y_1(t)+y_2(t)$ and taking into account the initial conditions 
$y_1(0)=1$, $y_2(0)=0$, we may solve ODEs (2) in the main text of the paper, obtaining
\begin{equation}  \label{z1}
z_1(t)\,=\, \exp\left[- \int_0^t \frac{\alpha}{x(s) (1-x(s))} \, ds\right] 
\end{equation}
and
\begin{equation}  \label{z2}
z_2(t)\,=\, 1 \,+\, \int_0^t \frac{\alpha(2x(s)-1)}{x(s) (1-x(s))}\, z_1(s) \, ds \;,
\end{equation}
where in Eq. (\ref{z2}), $z_1(s)$ is given by Eq. (\ref{z1}). 
The same path could be followed for the direct solution of the difference equations Eq. 
(1) in the main text, but formulae corresponding to Eqs. (\ref{z1}) and  (\ref{z2}) here 
are more involved and, more importantly,
the limit $N \rightarrow \infty$ will be appropriate for our 
further analysis. 
Of course Eqs. (\ref{z1}) and (\ref{z2}) may 
be trivially used to derive explicit expressions for $y_1$ and $y_2$, 
but we think the result is clearer in the form given by Eqs. (\ref{z1}) and (\ref{z2}). 

In general, $x(t)$ is a complicated function obtained by realizing the Wright-Fisher drift. 
In the $N \rightarrow \infty$ limit, it is a solution of the stochastic ODE
\begin{equation}
dx(t)\,=\, \sqrt{x(t)(1-x(t))} \, dw(t) \;,
\end{equation}
where $w(t)$ is standard Brownian motion, i.e. $E(dw(t))=0$ and $E((dw(t))^2)= dt$. 
As a consequence, we cannot explicitly compute the integrals in Eqs. (\ref{z1}) and (\ref{z2}). 
Anyway, Eqs. (\ref{z1}) and (\ref{z2}) can be used to give a qualitative description of the 
solutions to Eq. (2) in the main text and, if necessary, integrals may be easily numerically computed. 

As the integrand in the exponent of Eq. (\ref{z1}) is positive, 
it shows that the difference between $y_1$ and $y_2$ is positive and steadily decreasing. 
Moreover, this information, when plugged into Eqs. (2) shows that in fact $y_1$ 
decreases and $y_2$ increases. 

Eq. (\ref{z2}) on the other hand shows that gene flow from one subpopulation into the other 
is generally not symmetric. In fact, $z_2(t)-1= y_2(t)-(1-y_1(t))$ measures the difference 
between the fraction of type 1 alleles in subpopulation 2 and type 2 alleles in subpopulation 1. 
By Eq. (\ref{z2}), this difference decreases at times in which $x(t)<1/2$ and increases when 
$x(t)>1/2$. Moreover, it shows that gene flow is more effective at initial times, when $z_1(t)$ 
values are larger.

\section{Checking accuracy of the model and its stochastic simulation}
With the purpose of illustrating the qualitative behavior of the solutions of Eqs. 
(2), see appendix A, we show in Fig. S1 plots of $y_1(t)$ and $y_2(t)$ 
numerically obtained 
in the case of two deterministic histories $x(t)$ which illustrate typical situations 
occurring in the Wright-Fisher drift. 

It can be seen that all qualitative features of the solutions to Eqs. (2) are present. It should also be noticed in Fig. S1 that the final values of $y_1(t)$ and $y_2(t)$, i.e. their values at the time of extinction of one of the subpopulations, do depend very much on the history $x(t)$ and on the value of $\alpha$.

\begin{figure}
	\centering
		\includegraphics[width=0.9\textwidth]{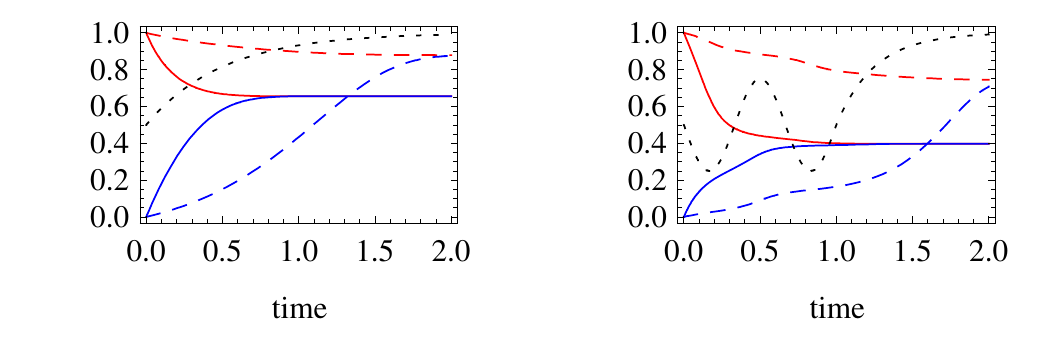}
			%\begin{picture}(21, 0.3)
		%\put(4.4,0){(a)}
		%\put(13,0){(b)}
		%\end{picture}
	\caption{\label{odeexamples} For two different histories $x(t)$ and two different values 
of $\alpha$ we plot the solutions of Eqs. (2). 
In both plots, the black dotted curve represents $x(t)$. 
The left plot corresponds to a situation in which subpopulation 2 is rapidly extinct, 
while the right plot to a situation in which extinction of population 2 occurs after an 
initial period of oscillating populations. In both pictures we represent a situation 
with $\alpha=1$ (full lines) and another with $\alpha=0.1$ (dashed lines). 
In each picture the upper (red) lines correspond to $y_1(t)$ and the lower (blue) 
lines to $y_2(t)$.
Notice that in these examples
the allelic fractions of the subpopulations become the same before extinction.}
\end{figure}

The final values of $y_1(t)$ and $y_2(t)$ are the 
most important outputs of the model, because they can be compared with experimental data. As stated above, these values are expected to heavily depend on the particular realization of $x(t)$ and on $\alpha$.  
Therefore, although the qualitative behavior of $y_1(t)$ and $y_2(t)$, as 
outlined in appendix A,
is quite well-understood, it is necessary to simulate the model by a computer program to 
obtain quantitative information on their final values.

We first simulate the history $x(t)$. This part begins by choosing a value for the 
total population size $N$ and a value $x(0)$ for the initial fraction of individuals 
in subpopulation 1. Some comments on the choice of $x(0)$ are made in appendix C. 
The choice of $N$ is not so relevant if it is large enough so that agreement between 
the solutions of Eqs. (1) and Eqs. (2) is good. In all results shown we 
have taken $N=100$, which produced a good agreement.

Then, individuals in generation $g=1$ independently and randomly choose the subpopulation to which they 
belong, being $x(0)$ the probability of choosing subpopulation 1. 
The fraction of individuals which expressed the choice for subpopulation 1 produces the value 
$x(1/N)$. In general, individuals in generation $g+1$ randomly and independently choose their subpopulation, being $x(g/N)$ the probability of choosing subpopulation 1. This procedure is repeated many times and it generates a realization of the 
Wright-Fisher drift, i.e. a sequence of 
values $x(k/N)$, $k=0,1,2, \dots$ until $x(t)$ attains either the value 0 or the value 1. 
If a realization of $x(t)$ is directly plugged into the difference 
equations (1), 
the \textit{theoretical} values of $y_1(t)$ and $y_2(t)$ can be easily obtained. 
These theoretical values have been used in producing e.g. results shown in Fig. 1, which represents the core of the paper.

The second part of the program concerns the processes of diploid 
reproduction and exchange of individuals between subpopulations. In order to obtain 
the \textit{simulated} values of $y_1(t)$ and $y_2(t)$, 
it is necessary to numerically run the stochastic processes of reproduction and individual exchange.
The former consists in the choice of the parents of each individual in the next generation 
and the latter in the random extraction of individuals to be exchanged between subpopulations. 
We will explain later the details of them. 
The purpose of this second part is twofold: it is necessary 
to check, see Fig. S1, the accuracy of the approximations made in deducing 
Eqs. (1), and also to obtain information concerning 
the common ancestors of all individuals in the population in paternal and maternal lineages. 
Although this information has no relevance for the
the final values of $y_1(t)$ and $y_2(t)$, it is necessary in order to
check whether or not the common ancestors of the whole 
population in paternal and maternal lines belong to the ancestors of the surviving subpopulation. 

In the second part of the program, at all time steps we suppose that half 
the number of individuals in any subpopulation are 
males and the other half females. 
The process of individuals exchange is simulated by randomly picking $\alpha$ 
individuals of subpopulation 1 and $\alpha$ individuals of subpopulation 2 and 
exchanging their subpopulation affiliation. 
In the more interesting case in which $\alpha$ is less than 1, 
we promote the exchange of 1 random individual of each subpopulation each 
$1/\alpha$ generations.

The reproduction process is simulated as follows:
each individual of subpopulation 1 at time $t+1/N$ makes a random choice of both his/her parents 
among male and female individuals in subpopulation 1 at time $t$, migrants included.
The analogous procedure is followed by individuals in subpopulation 2. 
At each generation we keep track of the entire genealogy of each of the $N$ individuals by 
counting the number of times each one of the ancestors
(individuals of the founding population which lived at time 0, before interbreeding started)
appears in his/her genealogical tree \cite{derrida1}. 
Then we proceed to computing the simulated value of the fractions $y_1(t)$ and $y_2(t)$.
We first consider a single individual at time $t$ in subpopulation 1 and we
count the number of times the ancestors belonging to subpopulation 1 appear in his genealogy,
then we divide this number by $2^{N t}$, and finally we average this value with respect to
all individuals of subpopulation 1 at time $t$.
The result is the \textit{simulated} value of $y_1(t)$. 
An analogous calculation produces the simulated value of $y_2(t)$. 

For each male individual at each generation we also keep track of his ancestor by paternal 
line in generation 0. For the female individuals we do the same for the maternal ancestor in 
generation 0. 

The left graph in Fig. S2 shows the result of one such simulation, 
in which we compare the theoretical and simulated values for $y_1(t)$ and $y_2(t)$ 
using the same Wright-Fisher drift history $x(t)$. It should be noted that, 
although not complete, agreement between simulated and theoretical 
quantities is good. We remind here that the simulated allelic fractions are subject to 
statistical fluctuations due to the random processes of exchange of 
individuals and diploid reproduction.

\begin{figure}
	\centering
		\includegraphics[width=0.9\textwidth]{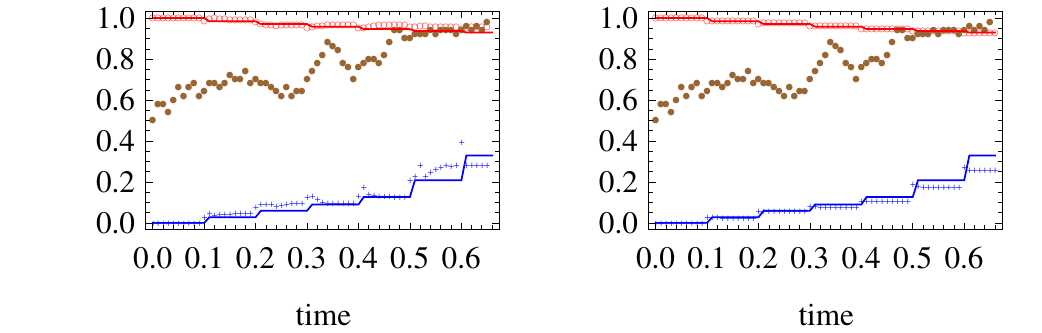}
			%\begin{picture}(21, 0.3)
		%\put(4.4,0){(a)}
		%\put(13,0){(b)}
		%\end{picture}	
	\caption{\label{oneandmany} For a single Wright-Fisher drift history $x(t)$ plotted 
with brown full dots and $\alpha=0.1$ we compare the theoretical and simulated values of $y_1(t)$ 
and $y_2(t)$. In both plots, the theoretical values are shown in full lines. The upper (red) line 
corresponds to $y_1(t)$ and the lower (blue) line corresponds to $y_2(t)$. 
The corresponding simulated values are shown respectively as red open dots and blue crosses. 
The left graph shows the simulated values obtained by a single simulation, whereas the right graph 
shows the averages of 100 simulations.}
\end{figure}

Indeed, we believe that the randomness in the diploid reproduction process accounts for the largest 
part of the difference between theoretical and simulated values. In fact, as shown in 
\cite{derrida1,derrida2,derrida3}, with diploid reproduction the contribution of each single 
individual to the gene pool some generations later is highly variable.  
On the other hand, if $\alpha$ is much less than $1/\log_{2}N$, randomness in the process of exchange of individuals is not so important
since at the time of exchanges the individuals in each subpopulation are already highly 
homogeneous from the point of view of allelic fractions.
We have directly checked this fact while producing the data shown in Fig. S2.
 
It should also be noticed that agreement between theoretical and simulated values is 
worst for $y_2$ when subpopulation 2 is close to extinction. 
In this case, in fact, given the small size of subpopulation 2, even a
small number of migrants induces large fluctuations in $y_2$.  

The right graph in Fig. S2 shows the average of the simulated values $y_1(t)$ 
and $y_2(t)$ over 100 simulations with the same history. Notice that the difference between 
theoretical and average simulated values is accordingly smaller.

\section{Why we must exclude values of $x(0)$ close to or in the experimental interval when inferring the value of $\alpha$}

As remarked in the main text, in producing Fig. 1, we have taken random values of $x(0)$ uniformly distributed between 0 and 0.8. The reason why we avoided larger values of $x(0)$ is that they are too close to the experimental interval $(0.96;0.99)$ or inside that interval. We now explain why this must be done.

First we observe that if $x(0)$ is in the experimental interval, 
then the final values of $y_1$ and $y_2$ will necessarily also lie in the experimental interval
provided that $\alpha$ is large enough. The free mating situation, in which subpopulations interact as if there were no differences among their members, is a particular case of this large $\alpha$ regime. Free mating, in the infinite population limit, is in fact ``described" by Eqs. (2) with an infinite value for $\alpha$. In this case the solution to the equations is straightforward: both $y_1(t)$ and $y_2(t)$ become instantaneously equal to $x(0)$.

The conclusion is that  if $x(0)$ 
lies in the experimental interval, then the model would fail to predict any upper bound to $\alpha$, as easy or free mating situations are allowed.
Nevertheless, we do not believe that either of 
these situations were likely to have occurred in reality, since 
distinct subpopulations coexisted for thousands of years.
Therefore, the experimental 
interval has to be excluded in the choice of $x(0)$.

\begin{figure}
	\centering
		\includegraphics[width=0.9\textwidth]{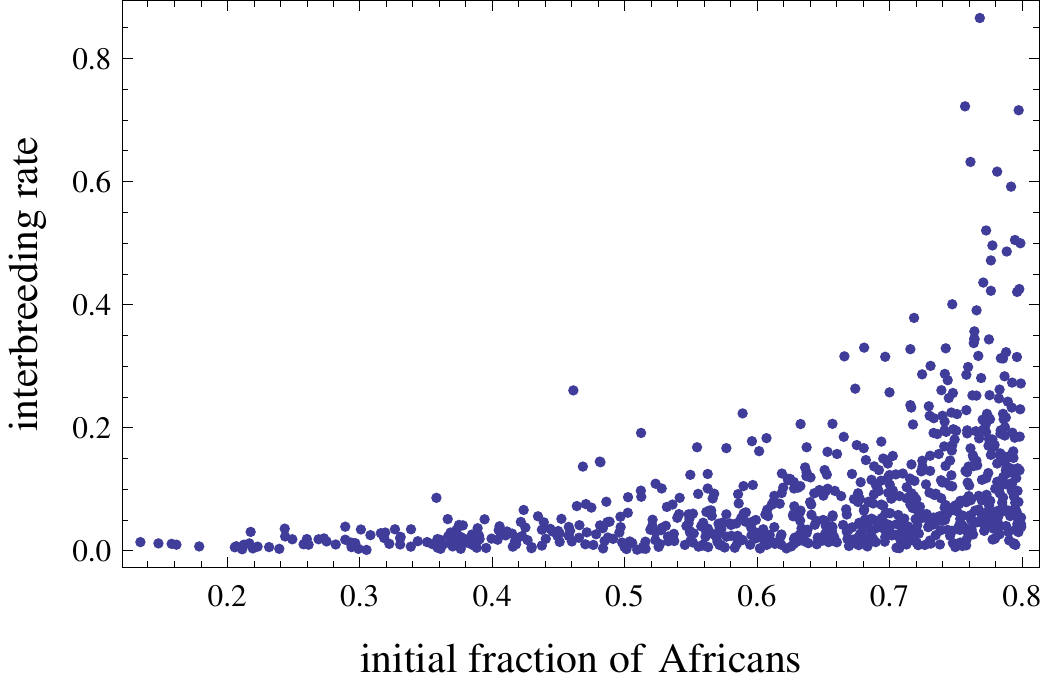}
			%\begin{picture}(21, 0.3)
		%\put(4.4,0){(a)}
		%\put(13,0){(b)}
		%\end{picture}
	\caption{\label{x0alpha} We plot here the correlation between $x(0)$ and $\alpha$ for 790 histories such that $y_1$ lies in the experimental interval. The histories were produced with random $\alpha$ and random $x(0)$ subject to $x(0) \leq 0.8$. Notice that the number of histories with $y_1$ in the experimental interval increases with $x(0)$, as well as the corresponding values of $\alpha$.}
\end{figure}

If we take instead values of $x(0)$ outside the experimental interval,
but still close to its boundaries, simulations show that both $\alpha_{\mathrm{mean}}$ 
and $\alpha_{\mathrm{max}}$ take very large values, such values tending to infinity as $x(0)$ gets closer to the experimental interval. This is illustrated in Fig. \ref{x0alpha}.
With $x(0)=0.8$ typical values of $\alpha$ become comparable to $1/\log_2 N$ (with the value $N=100$ 
we used and also with $N$ of the order of tens of thousands as it could have 
been in the real events in Middle East) or larger. 
As already commented, for such large values of $\alpha$, Eqs. (1) or (2) 
do not describe accurately the interbreeding process. 
The reason is that the assumption that all individuals in each subpopulation 
are genetically homogeneous, necessary to deduce Eqs. (1), fails.

Our choice in Fig. 1 of taking the values of $x(0)$ limited to 0.8 is thus a reasonable 
consequence of the mathematical characteristics of the model. It is also a reasonable choice 
from a historical point of view, because we are assuming that the Neanderthal subpopulation was
comparable to the African one; it might be smaller, but not extremely smaller, 
compared with the African one.

\section{Other results}
\begin{figure}
	\centering
		\includegraphics[width=0.9\textwidth]{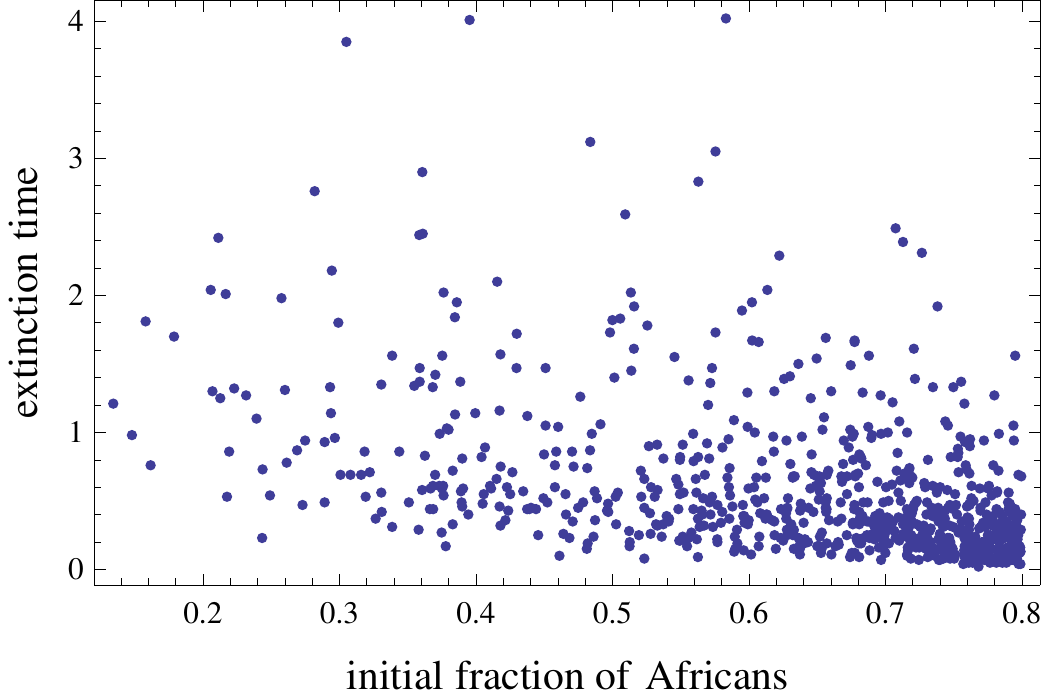}
			%\begin{picture}(21, 0.3)
		%\put(4.4,0){(a)}
		%\put(13,0){(b)}
		%\end{picture}
	\caption{\label{extinctimes} For the same 790 histories such that $y_1$ lies in the experimental interval as in Fig. \ref{x0alpha} we show the correlation between $x(0)$ and the time for Neanderthal extinction. The mean extinction time in the sample is 0.58.}
\end{figure}

The main result of our paper is the probability density distribution for the interbreeding rate $\alpha$ shown in Fig. 1. Other interesting resuts are illustrated here. In order to produce them we simulated a large number of Wright-Fisher drift histories $x(t)$ with random $x(0)$ constrained to be smaller than 0.8 and random $\alpha$ smaller than 2. For each history we numerically obtained the values of $y_1$ ad $y_2$ by iterating Eq. (1). We obtained then a set of 790 events such that the final value of $y_1$ lies in the experimental interval.

\begin{figure}
	\centering
		\includegraphics[width=0.9\textwidth]{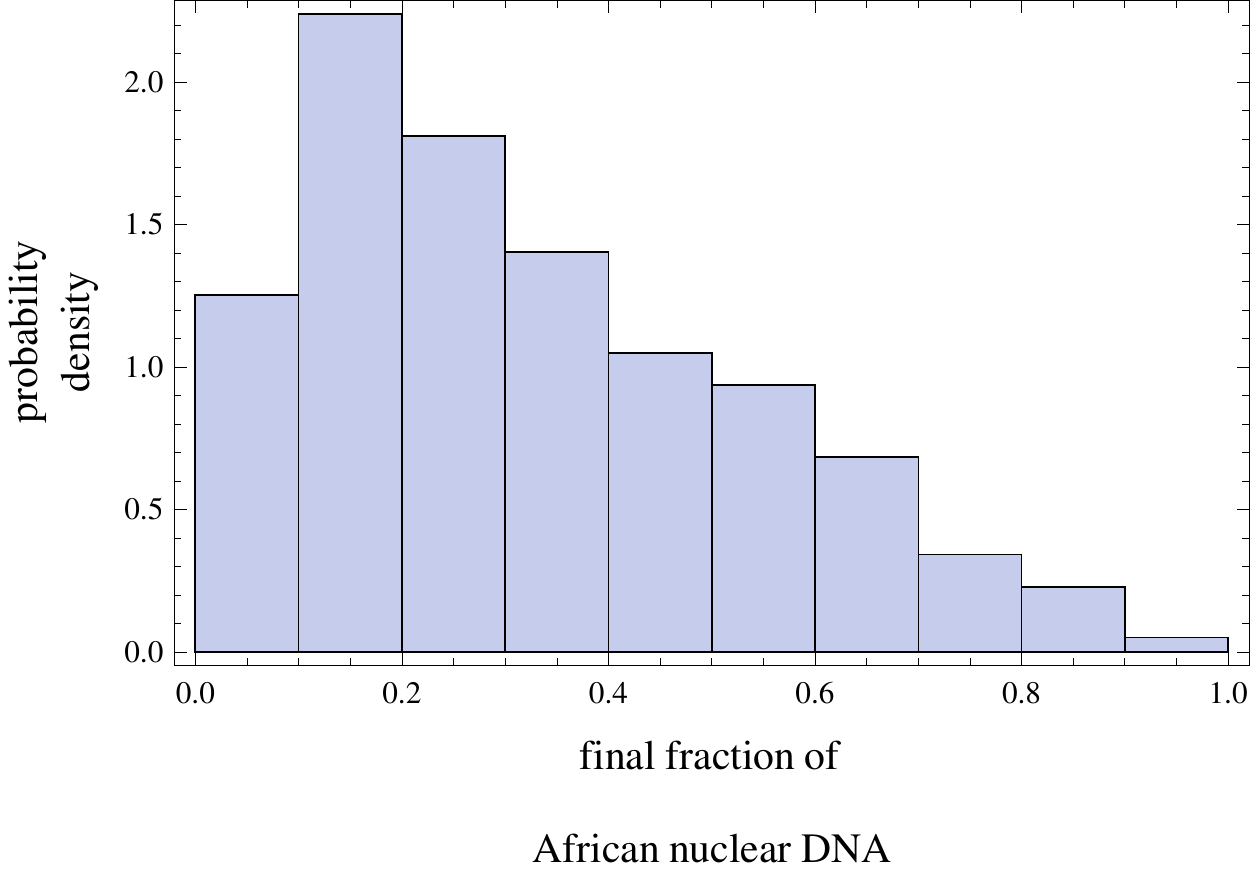}
			%\begin{picture}(21, 0.3)
		%\put(4.4,0){(a)}
		%\put(13,0){(b)}
		%\end{picture}
	\caption{\label{africanDNAneand} For the same 790 histories such that $y_1$ lies in the experimental interval as in Fig. \ref{x0alpha} we plot the probability density distribution for the final values of $y_2$. The mean final value of $y_2$ in the sample is 0.33.}
\end{figure}

Fig. \ref{x0alpha} was produced with the above mentioned sample. With the same sample we may also study the questions of extinction times, see Fig. \ref{extinctimes}, and the final values of $y_2$, i.e. how much African DNA was transmitted to Neanderthals before they were extinct in the Middle East, see Fig. \ref{africanDNAneand}. Comments on these figures were made in the main text.

\section{Mitochondrial DNA and Y chromosome}

Mitochondrial DNA and Y chromosome are both inherited in a haploid way. Furthermore mtDNA is 
not subject to recombination and recombination seems to be negligible for the Y chromosome. 
It is also believed that large portions of both are selectively neutral.
These facts allow an easier mathematical treatment of their statistical properties. 
From the experimental point of view, large samples of mtDNA \cite{cann} and Y chromosomes 
\cite{underhill} in living humans have been sequenced. 
The small variation among living humans is compatible with a single ancestor woman (mtDNA) and 
a single ancestor man (Y chromosome), probably both of African origin and living about 100-200 
thousand years ago. These facts have been interpreted as proofs of the Out of Africa model. 

More recently \cite{kringsetal}, the whole mtDNA of a few Neanderthal fossils became available. 
The average number of pairwise differences in mtDNA between a Neanderthal and a living human is 
significantly larger than the average number of pairwise differences in mtDNA among living humans. 
This has been considered as a further confirmation of the claim that Neanderthals belong to a 
separate species, see e.g. \cite{curratexcoffier}, and also for the Out of Africa model.

Both authors of this paper have separately claimed that the above facts are all compatible 
with anatomically modern Africans and Neanderthals being part of a single interbreeding 
population at the times they coexisted. 
In \cite{serva1}, using Kingman's coalescence, it was shown that 
the probability distribution of genealogical distances in a population 
of fixed size and haploid reproduction is random
even in the limit when the population size is infinite.  
The random distribution typically allows 
large genealogical distances among subpopulations.
In \cite{serva2} another important fact was statistically described: 
in a population of fixed size and haploid reproduction one of the two main 
subpopulations will become extinct at random times with exponential distribution. 
When such an extinction occurs, average genealogical distances 
among individuals in the population have a sudden drop. 
Finally, in \cite{serva3}, it was shown that mtDNA may be completely replaced in a population
by the mtDNA of another neighbor population, whereas some finite fraction of nuclear DNA persists.

These facts imply that the large genealogical distances 
between living humans and Neanderthals, as seen in mtDNA, 
are not uncommon in an interbreeding population. On the contrary, they turn out to be very 
likely if the correct statistics is used.
Furthermore, these facts imply that these distances may have been much larger at the time of 
Neanderthals extinction than they are nowadays. They also imply
that extinction of Neanderthals' mtDNA 
is compatible with the survival of their nuclear DNA. 

Exactly the same reasoning can be applied to the mitochondrial and nuclear DNAs 
of the fossil bones found in Siberia \cite{krauseetal, reichetal}, later described as the new population of Denisovans. The fact that Denisovans differ significantly both from Neanderthals and living humans in their mtDNA \cite{krauseetal} does not imply that they could not interbreed with either of them. Indeed, nuclear DNA proved \cite{reichetal} that they have interbred 
at least with some anatomically modern populations.

In \cite{nm1,nm2} the authors examined the question of survival of mtDNA and Y chromosome lineages 
in a population subject to exponential stochastic growth 
(supercritical Galton-Watson branching process). 
It was shown that exponential growth is compatible with the survival of a single 
mtDNA or Y chromosome lineage only if the growth rate is in a narrow interval. 
Thus, even if Neanderthals and anatomically modern Africans belonged to the same interbreeding 
population and even if this population was allowed to grow exponentially with a small rate, 
the more probable outcome would still be all humans being descendants either of a single woman 
(mtDNA) or a single man (Y chromosome).

In \cite{nm3}, the number of generations between successive branchings 
in the Galton-Watson process was computed. 
It was found that in the slightly supercritical regime, in which the survival of a single lineage 
is expected, trees typically have very long branches of the size of the whole tree along with 
shorter branches of all sizes. Thus, trees are qualitatively similar to those 
of the coalescent model and, as a consequence, the phenomenon of sudden drops in genealogical 
distances, described in \cite{serva2} is also present in this model.

\end{document}